# Wettability and Swelling Behavior of a Weak Polybasic Brush: Influence of Divalent Salts in the Environment

Chen Qu*, Zhongli Zheng

Department of Chemical Engineering, University of Notre Dame

## Abstract

We have studied the response of surface properties and swelling behaviors of annealed poly(2-vinyl pyridine) (P2VP) brushes covalently tethered to solid planar surfaces to divalent salts in aqueous solutions at varied pH values. Results derived from the quartz crystal microbalance technique, atomic force microscope and contact angle goniometry indicate that annealed polybase brushes undergo conformational transitions upon addition of divalent salts over a wide range of pH values below pKa: at low ionic strength, polybase brushes swell upon salts addition; at high ionic strength, polybase brushes collapse with salts addition. The extent and sensitive range of brushes conformational transition induced by divalent ions are found to be grater and broader than that caused by monovalent ions at similar ionic strength, indicating stronger effects on screening, osmotic pressure and bridging interaction. In addition, wetting measurements indicate that polybase-divalent counterions interactions can be used to switch surface characteristics from hydrophilic to hydrophobic in a predictable manner. The immediate implications of these experimental results are related to design of "smart" surfaces with controllable charge distribution, membrane thickness and wettability.

## Introduction

Organic thin films provide a versatile platform to introduce chemical functionalities that ultimately determine the overall characteristics of surfaces. An even more ambitious goal in material science is to design smart surfaces with controllable properties through accurate and predictable manner. Over the past few decades, polyelectrolyte brushes, which are monolayers of charged polymer chains with one end tethered to a substrate at sufficiently high grafting density, have attracted intensive attention for their sensitive response to environmental conditions like temperature,[1] pH,[2-7] and ionic strength.[8-16] The polyelectrolyte brushes with high tenability and stimuli-responsive “smartness” have shown to be extremely useful for many applications, such as colloidal stabilization,[17, 18] cell adhesion substrates,[19, 20] drug delivery,[21] biosensors,[22, 23] microreactors[24, 25] and microactuators.[26] A detailed understanding of the properties of such grafted polymer layers is therefore crucial if they are to be successfully employed in these proposed applications.

*Author to whom correspondence should be addressed. Electronic mail: cqu@nd.edu

For polyelectrolyte brushes, the influence of the environment on their charge is significant. An important distinction should be made regarding “quenched” and “annealed” polyelectrolyte brushes. For quenched (so-called “strong”) polyelectrolyte brushes, the position and number of charges along polymer chains are fixed; for annealed (so-called “weak”) polyelectrolyte brushes, the charge density depends on both pH and ionic strength of surrounding solution as a result of protonation-deprotonation equilibrium. The behavior of annealed polyelectrolyte brushes is principally controlled by electrostatic repulsion between charged monomers and the osmotic pressure of counterions. Consequently, in addition to pH, increasing ionic strength represents a second mechanism manipulating the structure and properties of annealed polyelectrolyte modified surface. The increase in ionic strength reduces the range and strength of repulsion between charges because of screening, while simultaneously increases the number of charged monomers along polymer chains.[27-30]

The swelling behavior of annealed polyelectrolyte brushes influenced by pH and ionic strength has been extensively studied. Different brush behavioral regimes exist depending on the chain length, grafting density, the charge fraction and the ionic strength of solution.[27, 28, 31] Theoretical studies predict that at sufficiently high salt concentration, the thickness of annealed polyelectrolyte brushes decreases with increasing monovalent ion concentration $c_s$ as a result of strong screening effect that significantly reduces the electrostatic repulsion between individual segments. In this regime the brush thickness H follows a scaling relationship $H \propto {c_s}^{-\frac{1}{3}}$ as that of quenched polyelectrolyte brushes.[27, 28] This regime is known as “salted brush regime” (SB). At low ionic strength, the counter ions concentration inside brushes layer is much higher than that of bulk solution, suggesting a high osmotic pressure exerting on brushes. The addition of salts leads to a substitution of counter ions (proton for polyacid or hydroxyl for polybase) by salt counterions. This results in an increase in charge fraction upon salts addition and the scaling relationship between brushes thickness and monovalent ion concentration is predicted to be

*Author to whom correspondence should be addressed. Electronic mail: cqu@nd.edu

$H \propto c_s^{\frac{1}{3}}$.[27, 28] This regime is called "osmotic brush regime" (OsB). In recent years, the swelling behavior of polyelectrolyte brushes has also been extensively investigated with a variety of characterization techniques, such as multi angle null ellipsometry,[5, 8, 11, 13] neutron reflection (NR),[5, 32] quartz crystal microbalance (QCM),[2, 9, 12, 16] atomic force microscope (AFM),[6, 7, 9, 12] X-ray photoelectron spectra (XPS)[7] and Fourier transform infrared spectroscopy (FTIR).[6, 9] Most experimental studies on annealed polyelectrolyte brushes focus on poly (acrylic acid) (PAA) and its derivatives.[13, 33-36] The predicted non-monotonic behavior of swelling as a function of monovalent ion concentration has been observed in polyacid systems, where at low and high ionic strength the brushes collapse.[13, 33-36] Recently, the non-monotonic response of a polybase, poly(2-(diethylamino)ethyl methacrylate) (PDEA), in the presence of potassium nitrate has also been reported.[8]

Although the effects of salts constituting of monovalent ions on the swelling behavior of annealed polyelectrolyte brushes have been intensively explored, the influences of divalent or multivalent ions on annealed polyelectrolyte brushes are relatively less studied. This is despite wide-spread importance in many situations of practical interest ranging from commercial products and processes, to environmental technologies, to biology. Rühe et al. studied the swelling behavior of poly(methacrylic acid) (PMAA) in contact with aqueous solutions containing alkaline-earth metal, copper and aluminum cations.[37] They found that complexes of chelate were formed via stable coordination bonding between cations and carboxylate anions in polymers and accordingly PMAA brushes were irreversibly collapsed. Zhulina et al. studied the swelling behavior of annealed polyelectrolyte brushes with multivalent ions through theoretical calculations and provided a quantitative description of scaling relationship.[38] To date, however, the interaction between annealed polybase brushes and divalent or multivalent counterions has never been experimentally investigated.

In the present study, we aim to investigate effects of divalent counterions on the surface and conformational properties of annealed polybase brushes. Poly(2-vinyl pyridine) (P2VP) brushes, which are widely used as model surfaces in protein adsorption,[39, 40] were prepared on planer solid surfaces through a "grafting to" route. The swelling behaviors and surface properties of such a system in aqueous solution were investigated as a function of the pH value of the solution and the concentrations of added divalent salts. The surface wettability of P2VP brushes was analyzed by comparing water contact angle at varied environmental -conditions. The hydration-dehydration behaviors of brushes were detected by QCM and further validated in combination with AFM. QCM is a very sensitive tool for the determination of adsorbed mass on flat substrates. Although it was firstly designed for monitor the rate of deposition in thin films in vacuum or air based on Sauerbrey equation,[41, 42] it has now become a popular tool for examining adsorption, conformation, and interactions of macromolecules in solution.[43-46] In addition to QCM, a direct AFM measurement of P2VP brushes thickness by liquid tapping mode AFM was utilized to obtain a complementary picture of structural changes of brushes. The results of these measurements were compared with theoretically predicted swelling behaviors of annealed polybase. The immediate implications of these experimental results are related to design of "smart" surfaces with controllable charge distribution, membrane thickness and wettability.

*Author to whom correspondence should be addressed. Electronic mail: cqu@nd.edu

# Experimental Section

### Materials

Single crystal silicon wafers with one side polished (Silicon Quest International) and microscope glass covers (Fisherbrand) were used as supporting surfaces of polymer brushes for AFM characterization and water contact angle measurements, respectively. The silicon wafers and glass covers were cleaned by first sonication in ethanol for 10 min followed by soaking in a heated piranha solution (30% $H_2O_2$ and 70% $H_2SO_4$) at temperature T=110 ℃ for 1h. Subsequently, the silicon wafers and glass covers were thoroughly rinsed with deionized water and dried with nitrogen (purity > 99.9%) before use. Polished, QCM crystals (SRS) with a Ti/Au coating and a fundamental resonant frequency of 5 MHz were used for QCM measurements. Amino end-terminated P2VP of 135,600 g/mol (PDI =1.04) were custom-synthesized by Polymer Source (Quebec, Canada). The molecular weights were characterized by gel permeation chromatography (GPC) method and reported by Polymer Source. 11-mercaptoundecanoic acid, sodium sulfate, sodium chloride and chloroform (Sigma Aldrich) were used without further purification. Acetone and ethanol of reagent grade (VWR) were used as received. Deionized water (Barnstead Nanopure Ⅱ) was used to prepare salt solutions.

### P2VP Brushes Synthesis

*Author to whom correspondence should be addressed. Electronic mail: cqu@nd.edu

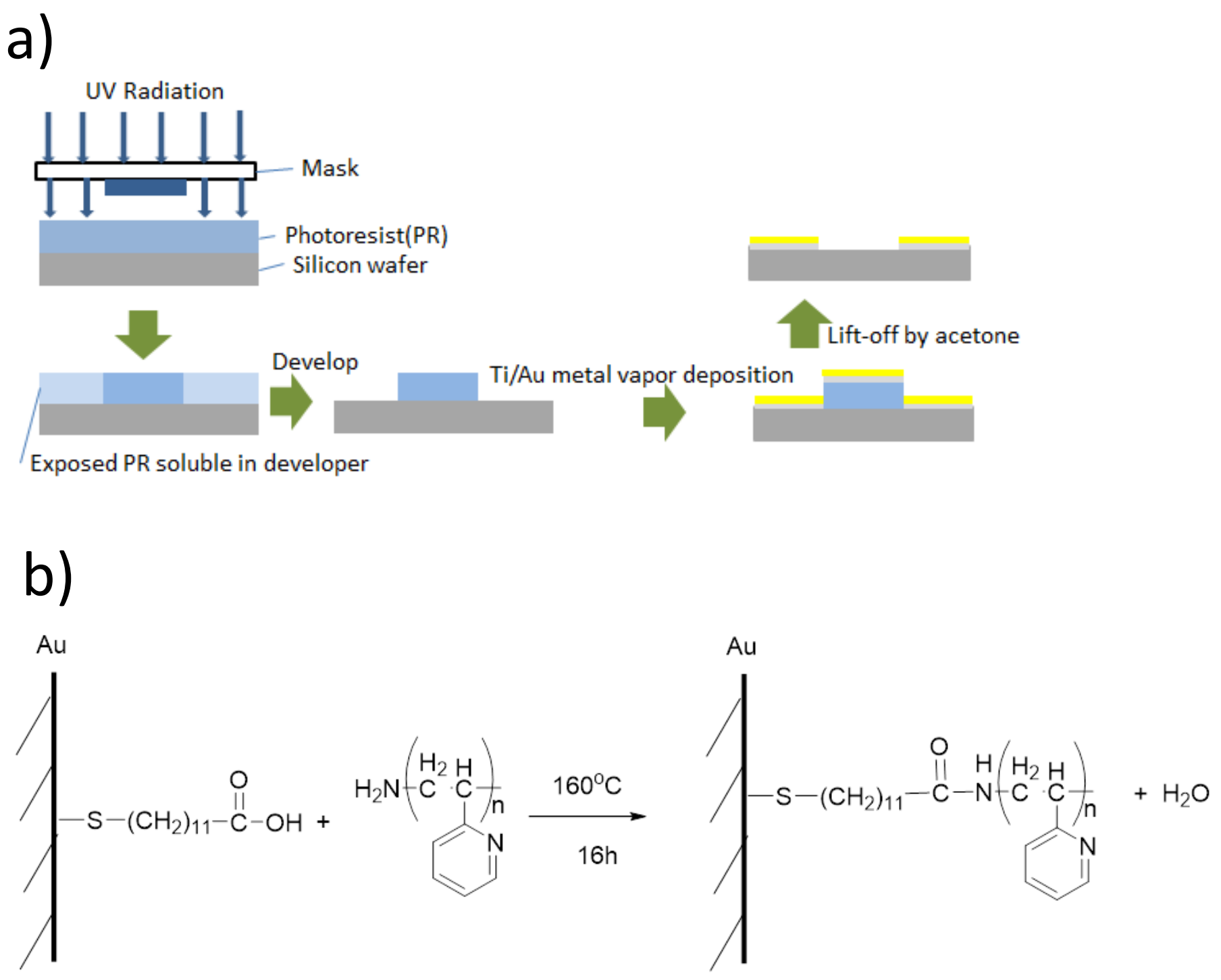

**Figure 1.** a) Outline of photolithography and metal deposition steps for fabrication of patterned PE brushes. b) The dehydration reaction between P2VP and 11-mercaptoundecanoic acid.

The preparation of P2VP brushes began with a Ti/Au metal deposition on supporting surfaces. Silicon wafers were cut into 1.3 × 1.3 cm pieces to fit the AFM sample holder. To get the brushes height from AFM cross-section image, a patterned Ti/Au surface was prepared through photolithography technique.[47] The procedure is shown in figure 1a. For glass coverslips, Ti/Au was directly deposited on the clean surfaces. The metal deposition was accomplished by the Oerlikon machine. Thickness of Ti and Au was 10nm and 30nm, respectively. The gold substrates were submitted to UV radiation for ten minutes and then submerged in ethanol solution of 20 μM 11-mercaptoundecanoic acid for a required time of 24 hours. The 11-mercaptoundecanoic acid was firmed adsorbed on gold surface through a strong thiol-gold binding to form a self-assembled monolayer (SAM) of coupling agents[48, 49] where the carboxylic acid group at the other end allowed for dehydration reaction with the amino terminated P2VP. The chemical equation is shown in figure 1b. P2VP was dissolved into chloroform with 1 wt%. Approximately 20 μL of the solution was placed on the SAM and spincoated at 4000 rpm for a minute and 8000 rpm for another 30 seconds. The newly made P2VP polymer brushes sample was then placed in a vacuum oven for 16 h at 160 ℃. Finally, the sample was removed from the oven and ultrasonicated twice for ten minutes in ethanol solution to remove unreacted polymers.

*Author to whom correspondence should be addressed. Electronic mail: cqu@nd.edu

## Contact Angle Measurements

Images of water droplets on the brushes surface were obtained using a contact angle goniometer (Rame-Hart, model 250). NaCl and Na2SO4 solutions under varied concentrations and pH values were prepared by resolving salts in aqueous stock solutions of different pH values. The pH of stock solutions was adjusted by varying the added amount of concentrated hydrochloric acid to deionized water and measured by a pH meter (Oakton pH6). Sessile contact angle titrations were performed no less than five sets of contact angles at each pH value for 4 μL drops.

## QCM measurements

The shifts in frequency ( $\Delta f$ ) for sensor crystal were measured using a commercial QCM (Stanford Research System, QCM200) consisting a chamber connected to the electronic unit. $\Delta f$ was determined for crystal coated with P2VP in deionized water. Solutions were prepared as mentioned previously and circulated over the coated surface of QCM crystals. While replacing the liquid present in the chamber, a large volume of new solution (~20ml) was circulated through the chamber to completely flush out the original liquid. Then the outlet of the chamber was closed to fill up with the solution of given salt concentration and pH. For each solution, the value of $\Delta f$ was allowed to equilibrate for at least 30 min.

## AFM Imaging

The P2VP brushes thickness was measured by taping mode atomic force microcopy (Nanoscope IV, Veeco). By AFM, the morphology of P2VP brushes was captured with a silicon probe (TESPAW, Bruker AFM Probes) in air and with a silicon nitride probe (NP, Bruker AFM Probes) in aqueous solution using a waterproof scanner (J scanner, Veeco). The solutions were prepared and circulated as previously described. For the AFM measurement conducted in aqueous solution, a tapping-mode fluid cell (MTFML, Veeco) with O-ring was used and cleaned with copious ethanol and blow-died in a stream of nitrogen before use. The setpoint amplitude was kept at lower limit while the trace and retrace overlapped each other well. To get high quality image in water, the resonance frequency of silicon nitride probe was fixed between 9 kHz and 10 kHz.

*Author to whom correspondence should be addressed. Electronic mail: cqu@nd.edu

# Results and discussion

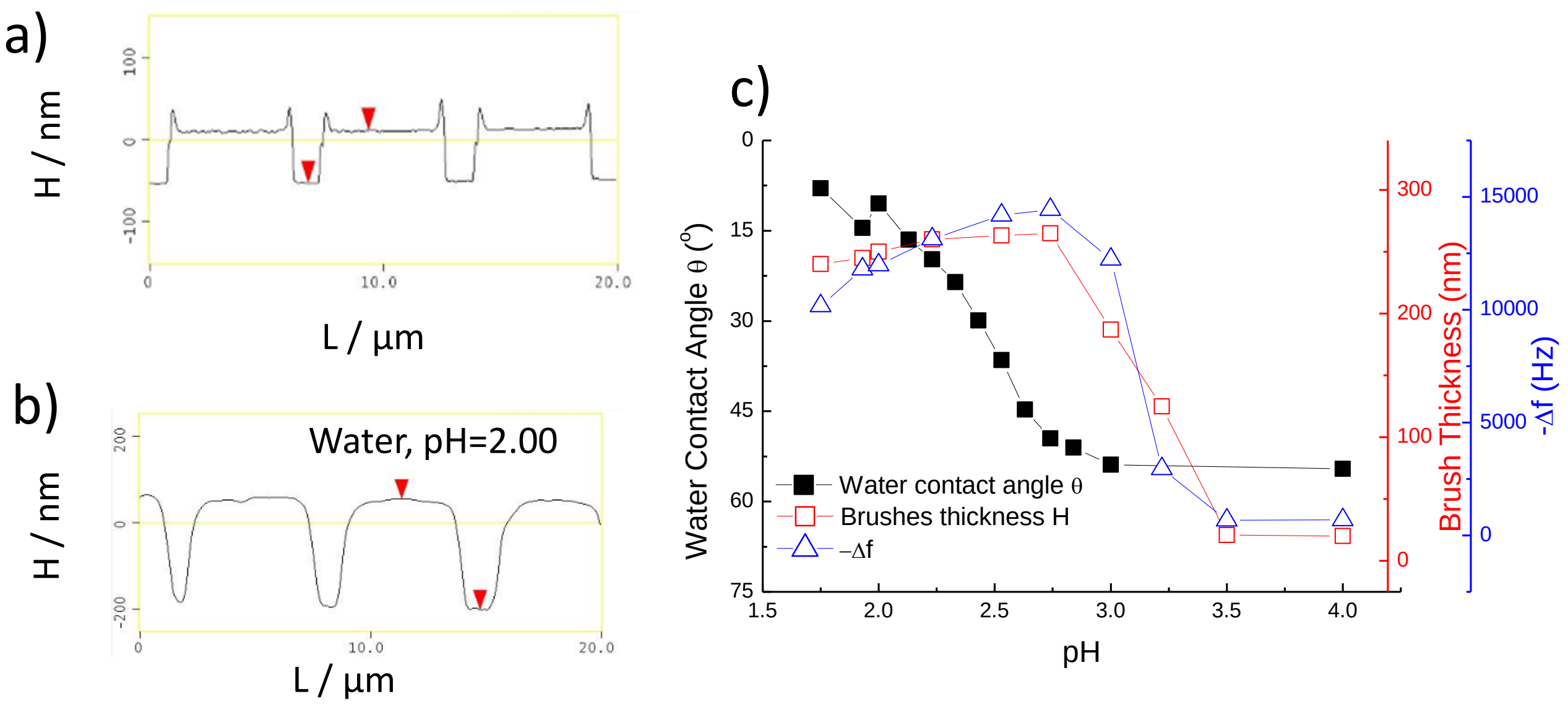

**Figure 2.** AFM cross-sectional analysis of patterned P2VP brushes: a) ~23-nm-thick brushes imaged in air and b) the same brushes imaged under water at pH=2.00. c) pH-dependence of P2VP brushes: water contact angle Θ (■), brushes thickness H from AFM (□) and changes in frequency – Δf from QCM ( Δ ).

### pH - Response of P2VP Brushes

The pH-responsive P2VP brushes were prepared through grafting-to method mentioned in experimental part. The dry P2VP brushes thickness was measured by tapping mode AFM, as shown in Figure 2a. Considering the step height includes both Ti-Au layer and P2VP brushes thickness, the P2VP brushes height is ~23nm. In the dry film condition, the brush can be assumed to be neutral since no counterions are present to induce charges in the chains. This assumption is the key to obtaining the grafting density, σ, as this equation for neutral brushes (eq 1) can be used.

$$\sigma = (H_{dry}\rho N_A) / M_n \qquad (1)$$

Where $H_{dry}$ is the dry brushes thickness, $N_A$ is Avogadro's number, ρ is the bulk PE density and $M_n$ is the molecular weight of polymers. The grafting density obtained in this case is ~0.10 chain/nm$^2$ and the distance between two anchoring sites is evaluated to be d ~3.2nm. Because $H_{dry} >> d$, the tethered P2VP chains form brushes.

*Author to whom correspondence should be addressed. Electronic mail: cqu@nd.edu

While P2VP brushes are generally believed to be neutral and adopt a totally collapsed conformation, the P2VP brushes can be significantly protonated in water at low pH, favoring a highly swollen conformation as a result of the strong repulsion between neighboring positive charges. This can be confirmed by an apparent increase in P2VP brushes thickness when immersed in acid water, as shown in Figure 2b. At pH=2.00, the P2VP brushes thickness is ~220nm, nearly ten times larger than that of dry state. In this case, P2VP brushes are highly charged and the increasing charge fraction results in a transition of surface properties from hydrophobic to hydrophilic.

To investigate the dependence of thickness, water uptake and the surface hydrophilicity of P2VP brushes on pH change, a pH-sweeping experiment was made with AFM, QCM and static contact angle goniometer, respectively. The results are shown in Figure 2c. At higher pH>3.5, the P2VP brushes are almost neutral and tend to adopt a collapsed conformation. The P2VP brushes thickness keeps constant at ~23nm, almost equal to that at dry state. When pH is decreased, P2VP brushes get partially protonated and the electrostatic repulsion between positive charges along P2VP chains forces brushes to adopt a swollen conformation, hence a gradual increase in brush thickness is observed from pH=3.5 to 2.7. Below pH=2.7, P2VP brushes are fully charged and the charge fraction doesn't increase further, therefore P2VP brushes thickness reaches a maximum value at ~220nm in this plateau. For QCM results, it is worth noting that $-\Delta f$ experiences a very similar trend vs pH with the P2VP brushes thickness measured by AFM. The interpretation of the $-\Delta f$ is somewhat complex. Growing $-\Delta f$ values are classically interpreted as an increase in adsorbed mass on the crystal surface, or in the case of a nonadsorbing solvent as a result of increased viscosity and/or density. Considering the mass of grafted P2VP chains are invariant and P2VP brushes layer is much thinner than the crystal, the change in $-\Delta f$ is primarily affected by coupled solvent mass within the P2VP brushes. When pH is decreased from 3.5 to 2.8, there is a significant rise in $-\Delta f$, indicating P2VP brushes are greatly hydrated with reducing pH, which is consistent with P2VP brushes thickness results. However, a small peak in $-\Delta f$ is observed at pH=2.8, where P2VP brushes at a totally swollen conformation begin to collapse with increasing pH. This non-monotonic behavior of $-\Delta f$ has been observed with some other environment-responsive polymer brushes at similar grafting densities.[50, 51] It is supposed that at very low pH, P2VP brushes are fully stretched, thus solvent molecules can freely go in and go outside P2VP brushes layer, resulting in a relatively poor performance at transmitting energy to surrounding fluids as crystal oscillates and lower effective coupled mass. However, with increasing pH, P2VP brushes start to collapse and the penetration of solvent molecules into and out of film is reduced. The result is an increase in effective coupled mass registered at surface. Additionally, P2VP brushes surface properties were studied by static water contact angle measurement. At pH>3.0, the water contact angle is nearly constant at 65 º, suggesting only a small portion of monomers are protonated and the charge fraction is low enough to exhibit a fairly hydrophobic behavior. However, at pH<3.0, water contact angle becomes smaller with increasing acidity, indicating a highly charged state and a hydrophilic surface. The transition ranges of water contact angle and P2VP brushes thickness are not exactly overlapping. This is understandable considering the two different types of characterization techniques.

*Author to whom correspondence should be addressed. Electronic mail: cqu@nd.edu

**Dependence of Water Contact Angle on Salt Concentrations**

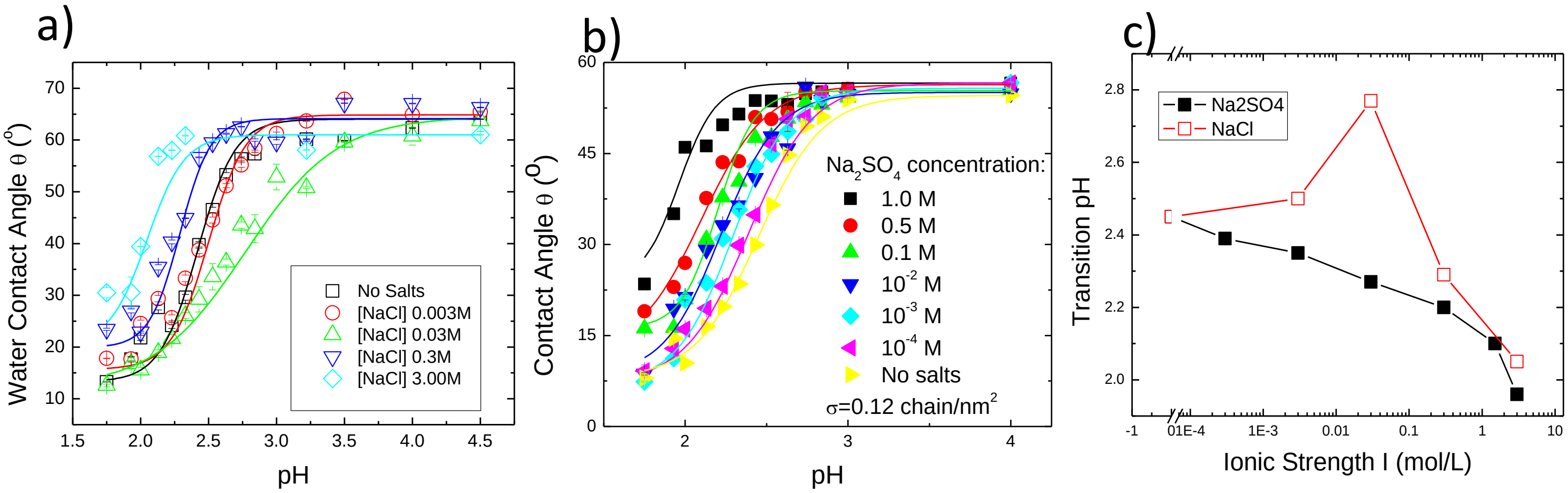

**Figure 3.** Water contact angle Θ of P2VP brushes depending on pH at different salt concentrations with a) NaCl and b) Na2SO4. c) Transition pH deduced from a) and b) as a function of ionic strength I with NaCl (☐) and Na2SO4 (■).

To investigate the influence of salts on surface properties of P2VP brushes, pH-sweeping water contact angles were measured under different salt concentrations. Firstly the monovalent salt NaCl was studied, as shown in Figure 3a. If we define the reflection point on the water contact angle vs pH curve as the transition pH implying a transition between hydrophobic and hydrophilic surface property, it would be 2.48 for P2VP brushes without any salts present. The pH-sweeping water contact angle curve shifts to higher pH when NaCl concentration is gradually increased from 0 to 0.03M, suggesting a more hydrophilic surface compared with bare P2VP brushes. However, pH-sweeping water contact angle curve doesn't shift to even higher pH with further increasing NaCl concentration, in contrast, the curve swifts to lower pH resulting in a more hydrophobic surface at high NaCl concentrations. At low NaCl concentrations, smaller water contact angles and higher transition pHs are resulted from higher charge fractions in P2VP brushes. P2VP brushes are in "Osmotic brushes region" (OsB) where ionic strength in bulk solution is weak and P2VP brushes conformation is dominated by high osmotic pressure exerted by counter ions within brushes layer. In OsB region, the charge fraction governed by chemical equilibrium in annealed PE brushes is increased as a result of counter ions re-distribution, therefore the surface become more hydrophilic and PE brushes are always expected to exhibit a more swollen conformation. However, the water contact angle curve shifts to opposite direction at high NaCl concentrations. This is because the P2VP brushes are in "salted brushes region" (SB) where the strong screening effect caused by high counter ions concentration weakens the repulsive electrostatic interaction between positive charges, resulting in a lower "apparent" or "effective" charge fraction.

*Author to whom correspondence should be addressed. Electronic mail: cqu@nd.edu

To investigate the influence of valence of counterions, a similar set of experiments with Na2SO4 were subsequently conducted. The results are shown in Figure 3b. The water contact angle remains nearly constant at ~65 $^{o}$ at pH>3.0 and reduces with increasing acidity. Different with NaCl, it is found that the water contact angle curve swifts to lower pH even at very low Na2SO4 concentrations, meaning a stronger screening effect induced by divalent counter ions compared with monovalent ones. P2VP brushes in SB region at pH<3.0 exhibit a more hydrophobic surface, suggesting a reduced "effective" charge fraction and probably a more collapsed conformation. However, it is difficult to tell the conformational change of P2VP brushes at pH>3.0, regarding the different responsive range of surface property and brushes thickness.

The influences of salt concentration on surface properties of P2VP brushes are summarized in Figure 3c. While the brushes surface become more hydrophilic at low NaCl concentrations and more hydrophobic at high NaCl concentrations, the surface always shows a more hydrophobic behavior in the presence of Na2SO4, indicating a stronger screening effect by Na2SO4.

**Dependence of Δf from QCM on Salt Concentrations**

*Author to whom correspondence should be addressed. Electronic mail: cqu@nd.edu

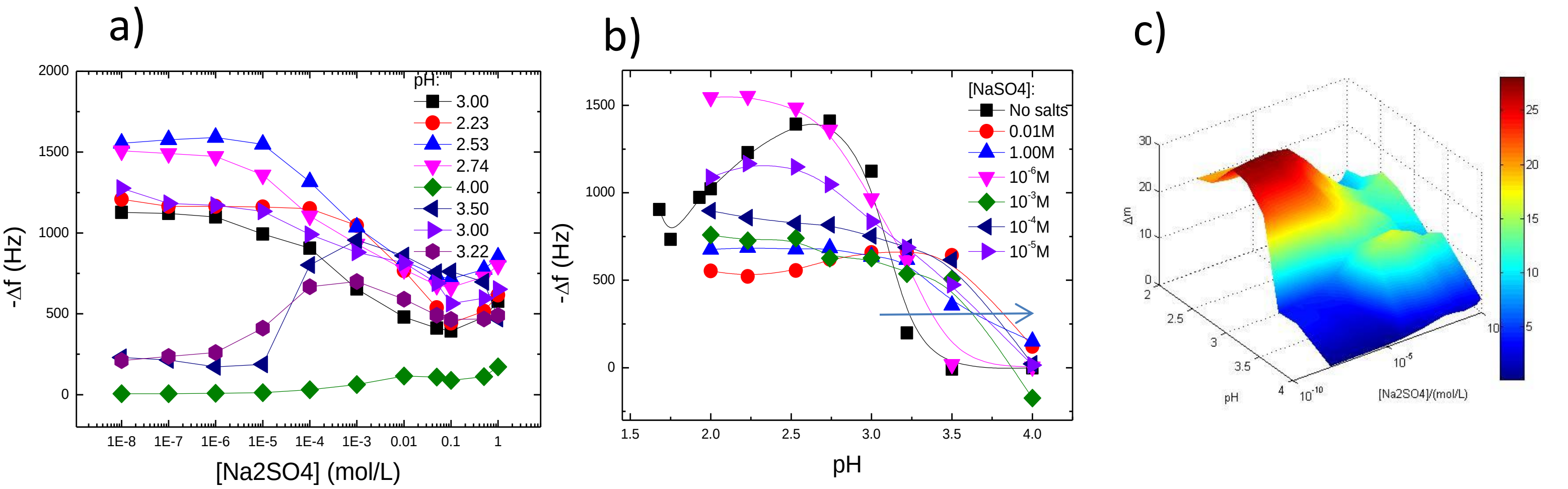

**Figure 4.** The changes in – Δf from QCM: a) depending on Na2SO4 concentration at different pH; b) depending on pH at different Na2SO4 concentrations and c) as a function of both Na2SO4 and pH in a three dimensional color-bar map.

QCM was employed to further explore the structural change of P2VP brushes. Figure 4a shows changes in – Δf as a function of Na2SO4 concentration at different pH. Each data points in the graph indicates the value when – Δf reaches equilibrium at the given salt concentration. Three regions can be distinguished according to different frequency-responses: for pH<3.00, – Δf decreases monotonically upon increasing Na2SO4 concentration; for pH>4.00, – Δf remains nearly zero at any Na2SO4 concentration and for 3.00<pH<4.00, – Δf exhibits a non-monotonic change upon Na2SO4 addition in which case – Δf firstly increases from very low – Δf to reach a maximum value and then goes down with further increasing salts concentration. The explanation of – Δf may be complex. A decrease in – Δf is generally interpreted as a loss of adsorbed mass on crystal surface, a decrease in non-adsorbing solvent coupled mass or a reduced solvent viscosity/density. In this case, the adsorbed mass of grafted P2VP brushes is invariant and the brushes layer is much thinner than the crystal, thereby changes in – Δf is mainly induced by coupled solvent mass within brushes layer. For pH<3.00, P2VP brushes are fully protonated and stretched. The HCl concentration in solution is higher than $10^{-3}$M and $Cl^-$ also serves as counter ions for screening. The – Δf decreases with increasing Na2SO4 concentration because P2VP brushes are in SB region and collapse upon salts addition. The collapse of annealed polyelectrolyte brushes caused by multivalent counter ions as a result of strong screening effect, cross-linking or pair-ions interaction has been reported by a few scientists in recently years.[15, 37,]

*Author to whom correspondence should be addressed. Electronic mail: cqu@nd.edu

[52] In these studies, the annealed PE brushes always experience an irreversible swollen-to-collapsed conformational transition with increasing multivalent counter ions amount and cannot recover from the collapsed state. For pH>4.00, the proton concentration in solution is too low for P2VP to get positively charged. The P2VP brushes are more like neutral brushes rather than polyelectrolyte brushes and adopt a totally collapsed conformation. In this case, P2VP brushes are in “neutral brushes region” (NB), being independent from salts in bulk solution.[27] It is worth noticing that for $3.00<pH<4.00$, – $\Delta f$ increases at low Na2SO4 concentration and then decreases at high Na2SO4 concentration. This indicates that P2VP brushes experience a gradual collapsed-to-swollen transition in low Na2SO4 concentration range, and then collapsed in the presence of large amount of Na2SO4. The re-swollen behavior happens in OsB region, where the degree of protonation in P2VP brushes is increased by counter ions redistribution and consequently the electrostatic repulsion between neighboring charges is enhanced. This re-swollen behavior for annealed polyelectrolyte brushes in OsB region caused by multivalent counter ions has been predicted in a few theoretical studies.[38] However, to the best of our knowledge, this is the first time for this behavior to be confirmed experimentally. The peak in – $\Delta f$ curve obtained from QCM provides the first experimental evidence of this re-swollen phenomenon. With further increasing Na2SO4 concentration, P2VP brushes enter SB region and start to collapse, resulting in reduced water content within brushes layer.

To further confirm P2VP brushes structural changes influenced by salts concentration and pH, a subsequent set of QCM experiments were made by increasing pH at different Na2SO4 concentrations, as shown in figure 4b. It clearly shows that at any given Na2SO4 concentration, P2VP brushes collapse with elevated pH. The initial water content inside brushes at pH=2.00 is determined by Na2SO4 concentration, in which case higher Na2SO4 concentration indicates less water content. When pH is increased, P2VP brushes transform from swollen/partial-swollen state to collapsed state. The conformational transition pH shifts to higher pH values with increasing Na2SO4 concentration, which is opposite to the water contact angle transition pH. The contrary trends are not difficult to understand: while the water contact angle is essentially dominated by “apparent” charge fraction that is lowered by Na2SO4 screening, the brushes conformation is a comprehensive result of both the degree of protonation and the counter ions screening, where the “real” charge fraction has been increased to maximum in OsB region and thereby more salts are required to sufficiently screen positive charges to convert brushes to collapsed state.

*Author to whom correspondence should be addressed. Electronic mail: cqu@nd.edu

If we compare – Δf changes in figure 4a and 4b, it is evident that – Δf values in these two plots show significant consistency with each other, meaning the extent of swollen of P2VP brushes can be determined by a given set of pH and Na2SO4 concentration, regardless detailed experimental history. Considering the hydration and dehydration should have dominant effect on – Δf, the apparent coupled solvent mass Δm inside P2VP brushes can be estimated using Sauerbrey equation (eq 2):[41]

$$\Delta m = -\frac{\Delta f}{C} \quad (2)$$

With the mass sensitivity constant, C=56.6 ng/cm$^2$. This can at least give a qualitatively correct picture of hydrodynamic behavior of P2VP brushes. Figure 4c shows a three dimensional map of Δm as a function of both pH and Na2SO4. OsB region, SB region and NB region discussed previously are clearly represented. The intuitive presentation of – Δf implies that there is a possibility of creation of smart coating films at desired thickness, permittivity and rigidity via tuning environment conditions.

**Dependence of P2VP Brushes Height on Salts Concentrations**

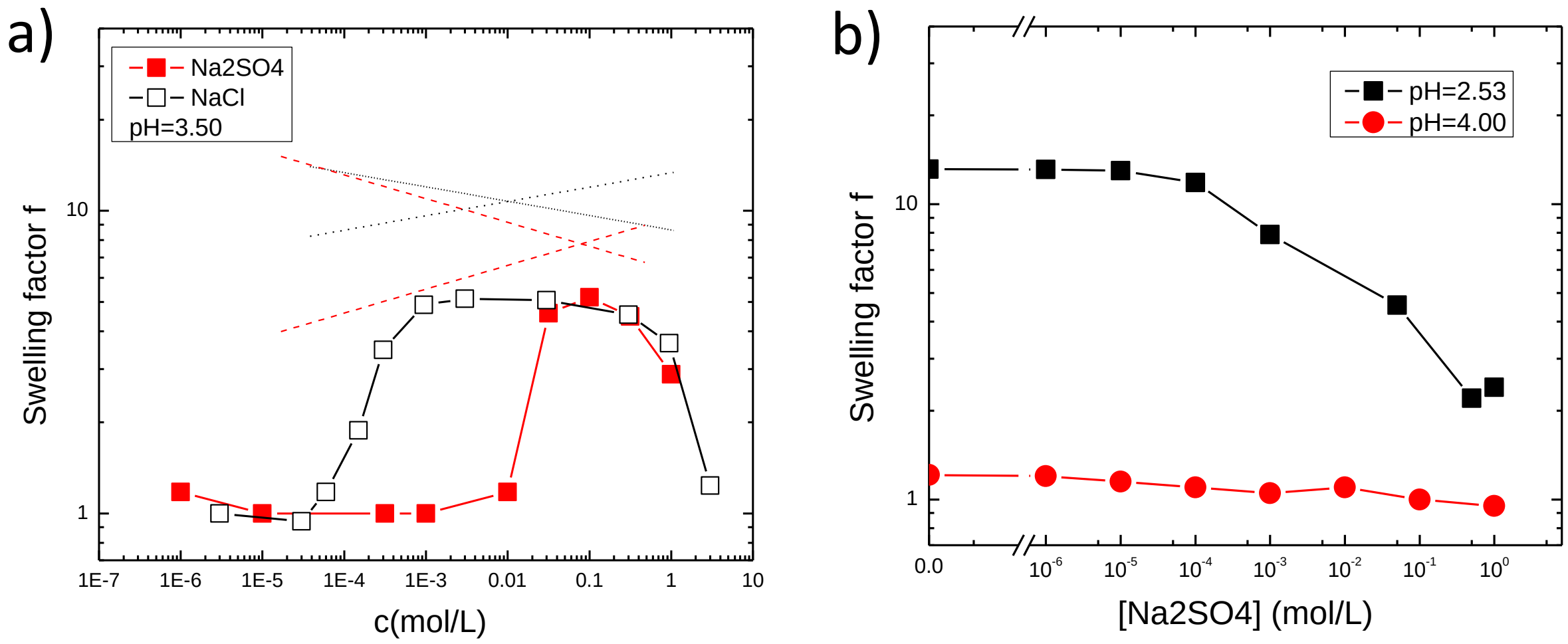

**Figure 5. a)** Swelling factor of P2VP brushes as a function of the concentration of NaCl (□) and Na2SO4 (■) at pH=3.50. The dashed line and dotted line represent the theoretically predicted scaling behavior for NaCl and Na2SO4 respectively. b) Swelling factor of P2VP brushes as a function of Na2SO4 concentration at pH=2.53 (■) and pH=4.00 (●).

To further investigate the swelling behavior of P2VP brushes and compare the results with scaling laws, tapping mode AFM was employed to give quantitative information of brushes conformation. A swelling factor f is defined as the ratio of the thickness of swollen brushes H to that of dry brushes $H_{dry}$ for better description of the extent of swelling. Figure 5a displays the swelling factor of P2VP brushes as a function of added NaCl or Na2SO4 concentration at pH=3.50. For comparison, the theoretically predicted scaling behavior for weak PE brushes in the presence of monovalent or divalent counter ions are shown in the same graph as well.[38] In both cases of NaCl and Na2SO4, P2VP brushes height first increases with increasing salt concentration in OsB regime and then decreases in SB regime. This is well understood on the basis of a shift in the protonation/deprotonation equilibrium at low salt concentration, which leads to the generation of more charges on the chains and thus a stronger chain stretching, and then the screening of charges at high salt concentrations, which leads to a decrease of the electrostatic repulsion. The non-monotonic behavior of annealed PE brushes caused by monovalent counter ions has also been observed in planer poly(methacrylic acid) brushes (PMAA) by Rühe et al.,[13] poly(acrylic acid) (PAA) brushes by Currie et al.[33] and poly(2-(diethylamino)ethyl methacrylate) (PDMA) brushes by Wanless et al.[8] However, no experimental report has been published so far on the re-swollen behavior induced by divalent counter ions with PE brushes.

*Author to whom correspondence should be addressed. Electronic mail: cqu@nd.edu

It is evident that P2VP brushes start to swell upon Na2SO4 addition at a much lower concentration limit ($10^{-4}$M) than NaCl (0.01M), showing a broader OsB regime. This demonstrates a greater influence brought by divalent $SO_4^{2-}$ on ion-redistribution near positively charged P2VP chains. The physical picture of this should originate from the intrinsic difference of these two couterions, as discussed below qualitatively. In the first case, the divalent $SO_4^{2-}$ result in a greater contribution to the increase of osmotic pressure within brushes layer when the original conterions $OH^-$ are substituted. The quickly growing osmotic pressure accordingly promotes a more extended conformation compared with that induced by monovalent cunterions. Also, it is thought that $SO_4^{2-}$ can interfere with the hydrogen bonding between amides or ammoniums in the polymer and water molecules, which facilitates the penetration of $SO_4^{2-}$ from bulk solution into brushes layer and leads to a more rigid brushes layer.[53] A third factor could be the bridging effect. Considering that $SO_4^{2-}$ with larger size and higher valence can possibly serve as cross-link points between charged segments along polymer chains, the bridging effect consequently results in the formation of polyelectrolyte-sulfate complex system that is mechanically more stable and easier to detect. In SB regime, the brushes height decreases steeper with Na2SO4, probably resulting from a stronger screening by divalent counter ions and special cross-linking interactions determined by chemical natures of different ions. The deviation of swelling factors of P2VP brushes between NaCl and Na2SO4 case is in good agreement with theoretical prediction reported by Zhulina et al qualitatively.[38] But it is obvious that the exponents both in OsB and SB regimes are much larger than the exponents from theoretical calculations (f $\propto c_s^{\frac{1}{3}}$ for NaCl and f $\propto c_s^{\frac{1}{5}}$ for Na2SO4 in OsB regime, f $\propto c_s^{-\frac{1}{3}}$ for NaCl and f $\propto c_s^{-\frac{1}{5}}$ for Na2SO4 in SB regime). This might be due to the influence of polydispersity of chains, the hydrophobicity of the polymer backbone, the nontrivial three-body interactions caused by high grafting density of the system, or the special bridging interaction between divalent ions and charged polymer chains, which were not taken into consideration in theoretical calculations. Also, the swelling factor of P2VP brushes in presence of Na2SO4 shows a broad plateau in the crossover between OsB and SB regime. This is probably because the swelling and collapsing effects coexist and compete with each other in the brush system and accordingly a plane of wide Na2SO4 concentration is observed. It is worth noticing that in spite of being partially-swollen with Na2SO4 or NaCl, P2VP brushes surface remains fairly hydrophobic. The particular combination of brushes thickness and wettability property enables P2VP brushes to serve as a good candidacy for barrier on coating to stop or change the speed of a fluid in nano-device.

*Author to whom correspondence should be addressed. Electronic mail: cqu@nd.edu

Figure 5b displays the swelling factor as a function of Na2SO4 concentration at pH=2.53 and pH=4.00. As discussed previously, P2VP brushes are fully protonated at pH=2.53 in SB regime and collapse with further addition of Na2SO4 as a result of screening. The swelling factor drops from an original value of ~14 to ~2 when the Na2SO4 concentration is gradually increased to 1.0M. At pH=4.00, P2VP brushes are more like neutral brushes and show little dependence on ionic strength in bulk solution. The swelling factor remains nearly 1 in the whole range of Na2SO4 concentration. The results from tapping mode AFM provide valid quantitative information about the extent of P2VP brushes swelling, showing great consistency with QCM results earlier described.

## Conclusions

In summary, the surface properties and swelling behavior of P2VP brushes grown on a solid surface through grafting-to process was studied using water contact angle measurement, QCM and tapping mode AFM. The samples were exposed to a variety of solutions containing low molecular weight divalent counter ions at different concentrations. P2VP brushes exhibit conformational change over a broad pH range below the pKa of brushes: at low ionic strength, P2VP brushes swell upon divalent salts addition; at high ionic strength, P2VP brushes collapsed with further increasing divalent salts concentration. The ion-sensitive range and conformational change with divalent counterions is found to be broader and greater than that with monovalent ions. Additionally, wettability results demonstrate that unlike monovalent salts that turn brushes surface more hydrophilic at low concentrations, divalent conterions always lead to a more hydrophobic surface in a wide range of concentration. The different influences of divalent counterions compared to monovalent ions are possibly resulted from stronger screening, bridging and osmotic pressure effect. The experimental results provide immediate reference for the fabrication of "smart" thin films with desired thickness and hydrophilicity.

*Author to whom correspondence should be addressed. Electronic mail: cqu@nd.edu

*Author to whom correspondence should be addressed. Electronic mail: cqu@nd.edu

*Author to whom correspondence should be addressed. Electronic mail: cqu@nd.edu

*Author to whom correspondence should be addressed. Electronic mail: cqu@nd.edu

*Author to whom correspondence should be addressed. Electronic mail: cqu@nd.edu